\documentclass[twocolumn]{aastex701}

\begin{document}

\title{Survival or Destruction: Effects of Spheroidal Satellite Collisions on Bars in Milky Way-Like Galaxies}

\correspondingauthor{Y. F. Zhou \& Z. Li}

\author[0009-0006-0194-6211]{Yufan Fane Zhou}
\email[show]{yufanz@smail.nju.edu.cn}
\affiliation{School of Astronomy and Space Science, Nanjing University, Nanjing 210046, China}
\affiliation{Key Laboratory of Modern Astronomy and Astrophysics (Nanjing University), Ministry of Education, Nanjing 210046, China}

\author[0000-0003-0355-6437]{Zhiyuan Li}
\email[show]{lizy@nju.edu.cn}
\affiliation{School of Astronomy and Space Science, Nanjing University, Nanjing 210046, China}
\affiliation{Key Laboratory of Modern Astronomy and Astrophysics (Nanjing University), Ministry of Education, Nanjing 210046, China}
\affiliation{Institute of Science and Technology for Deep Space Exploration, Suzhou Campus, Nanjing University, Suzhou 215163, China}

\author[0000-0001-7434-5165]{\'Oscar Jim\'enez-Arranz}
\email{oscar.jimenez_arranz@fysik.lu.se}
\affiliation{Lund Observatory, Division of Astrophysics, Department of Physics, Lund University, Box 43, SE-22100 Lund, Sweden}

\author[0000-0002-6299-152X]{Santi Roca-F\`abrega}
\email{santi.roca_fabrega@fysik.lu.se}
\affiliation{Lund Observatory, Division of Astrophysics, Department of Physics, Lund University, Box 43, SE-22100 Lund, Sweden}
\affiliation{Departamento de F\'isica de la Tierra y Astrof\'isica, Fac. de C.C. F\'isicas, Universidad Complutense de Madrid, E-28040 Madrid, Spain}

\begin{abstract}
Although stellar bars are prevalent in local galaxies, unbarred galaxies constitute a significant fraction, particularly at high redshifts. While some galaxies are unbarred by nature due to stability against the bar instability, several mechanisms capable of transforming barred galaxies into unbarred systems have also been proposed, such as central mass concentration, specific dark matter halo morphologies and tidal interactions. Regarding galactic interactions, mergers can undoubtedly disrupt bars while potentially destroying the entire disk. However, the effects of pure collisions (non-merging scenarios) on bars remain unclear, with limited existing studies yielding contradictory conclusions. Here we aim to systematically investigate the disruptive effects of collisions on bars hosted by Milky Way-like galaxies using \textit{N}-body/SPH simulations. We model collisions between the barred galaxy and a spherical intruder, conducting multiple simulations by varying interaction parameters, with mass ratios set at 1:3, 1:5, and 1:15. We find that bars are remarkably robust, with most interactions failing to significantly reduce their strength or pattern speed. Only off-center high-inclination retrograde collisions can effectively destroy bars, while central high-inclination collisions can substantially decrease the pattern speed. Such destruction and deceleration primarily arise from gravitational forces rather than gas-related processes. Notably, compared to collisions occurring along the bar's major axis, those along the minor axis cause greater weakening but can slow the bar's natural deceleration. Furthermore, changes in mass resolution do not significantly affect the results when the resolution is better than $\sim$$10^{5}$\,${M}_{\odot}$.
\end{abstract}



\section{Introduction}
Stellar bars are a kind of non-axisymmetric structures in the inner parts of disk galaxies. They play a crucial role in galaxy evolution by altering the mass distribution and morphological properties of galaxies \citep[e.g.,][]{sakamoto1999,athanassoula2002,kormendy2004}, and also influencing the star formation efficiency and the AGN activity \citep[e.g.,][]{ho1997,jogee2005,hopkins2010,galloway2015}.

Based on optical and infrared observations, a significant number of disk galaxies in the local universe exhibit bars \citep{eskridge2000,sheth2008,cheung2013,buta2015}, while bars are also present but show a declining fraction at high redshifts \citep{guo2023,leconte2024,guo2025,geron2025}. Given that numerical simulations suggest bars can readily form through self-gravitating instabilities of disks  \citep{efstathiou1982,roca2013,bland2023} or tidal interactions with perturbers \citep[e.g.,][]{gerin1990,lokas2018,cavanagh2020,jimenez2024,jimenez2025,zhou2025}, unbarred local galaxies have drawn significant attention \citep{saha2018}, with certain cases such as M33 remaining inadequately explained \citep{sellwood2019}.

Substantial efforts have been devoted to explaining the destruction of stellar bars. In contrast to simulations considering only stellar components, \citet{hasan1990,pfenniger1990,norman1996,hozumi2005} demonstrated that the presence of supermassive black holes or central mass concentration fed by gas inflow can dissolve bars, though \citet{shen2004,athanassoula2005} noted that the required central mass may need to be unreasonably large for this mechanism to be effective. \citet{el-zant2002} suggested that bars could also be dissolved if they are embedded in a triaxial and centrally concentrated dark matter halo. Furthermore, the dual role of tidal interactions has been recognized: although they can induce bar formation as mentioned earlier, they may also exert disruptive effects on bars \citep[e.g.,][]{athanassoula1996,nishida1996}.

Simulations by \citet{sellwood1996,athanassoula1999}, which studied satellite galaxies colliding with barred galaxies on rectilinear trajectories, suggest that if the perturber's mass is small, it fails to disrupt the bar; if sufficiently large, it destroys the entire disk instead. Thus, \citet{athanassoula1999} also initialized the companion on a nearly circular orbit, and the merger successfully destroyed the target galaxy's bar, which is a scenario also frequently occurred in cosmological simulations \citep[e.g.,][]{cavanagh2022,lu2025}. A limited number of non-merging interaction simulations still achieved success. For example, using 2-dimensional \textit{N}-body simulations, \citet{nishida1996} demonstrated that an off-center vertical collision whose impact point lies along the bar's minor axis can destroy the bar, whereas bars are not destroyed if the collision happens along its major axis or at the center. However, \citet{berentzen2003} reached a contradictory conclusion using \textit{N}-body/SPH simulations incorporating gas components, demonstrating that central collisions cause greater damage to bars than peripheral ones. Additionally, in KRATOS \citep{jimenez2024}, a suite of simulations focusing on the Large and Small Magellanic Clouds system, some collision scenarios successfully disrupt bars, though it remains unclear which types of collisions are consistently effective.

In summary, the destructive effects of tidal interactions, particularly in non-merging scenarios, on stellar bars remain an area requiring further investigation. It should be noted that disruption of bars by interaction is not merely a phenomenon in numerical simulations, but also supported by observational evidences. The statistical study of \citet{lee2012} on the correlation between bar occurrence rates and galactic environments indicates that strong tidal interactions can destroy bars. Observations of galaxy pairs also reveal that bar fraction decreases with reduced pair separation or increased companion mass, suggesting the disruptive capability of tidal interactions \citep[e.g.,][]{casteels2013,tawfeek2024,li2026}.

Therefore, here we aim to systematically investigate bar destruction through collisions (non-merging scenarios), thereby filling a gap in existing researches. Following the methodology of \citet{zhou2025} on collision-induced bar formation, this work explores the parameter space of interactions to study the inverse problem, collision-induced bar destruction. We present multiple sets of \textit{N}-body/SPH simulations involving an initially barred Milky Way (MW)-like galaxy and a spherical intruder, monitoring the bar's evolution after their collision. Intruder mass, relative velocity, impact parameter and inclination angle are varied to determine what kinds of collisions can destroy galactic bars. The fiducial model of the target galaxy incorporates gas components, while comparative simulations without gas (i.e. pure \textit{N}-body) are also conducted.

The rest of the paper is structured as follows. Section~\ref{sec:methods} describes the parameters of the target galaxy and the intruder, the configuration of collisions, and the Fourier decomposition method used for bar analysis. The main results of our simulations are presented in Section~\ref{sec:results}, while Section~\ref{sec:discussions} offers some additional results and discussions. A brief summary is given in Section~\ref{sec:summary}.

\section{Methods}
\label{sec:methods}

\subsection{Galaxy models}
\label{subsec:models}
We generate the initial condition of our target galaxy using the methodology described in \citet{springel2005ic}. The target galaxy comprises a Navarro–Frenk–White \citep[NFW;][]{navarro1996} dark matter halo, an exponential disk with stellar (70\% of the disk mass) and gaseous (30\% of the disk mass) components, and a Hernquist \citep{hernquist1990} stellar bulge, with parameters referenced from the model of the Milky Way in \citet{klypin2002} as detailed in Table~\ref{tab:1}. The value of Toomre parameter is $Q = 0.9$. The mass resolutions of dark matter and baryons are $2\times10^{5}$\,${M}_{\odot}$ and $1\times10^{4}$\,${M}_{\odot}$, respectively. 

\begin{figure*}
    \centering
    \includegraphics[width=\linewidth]{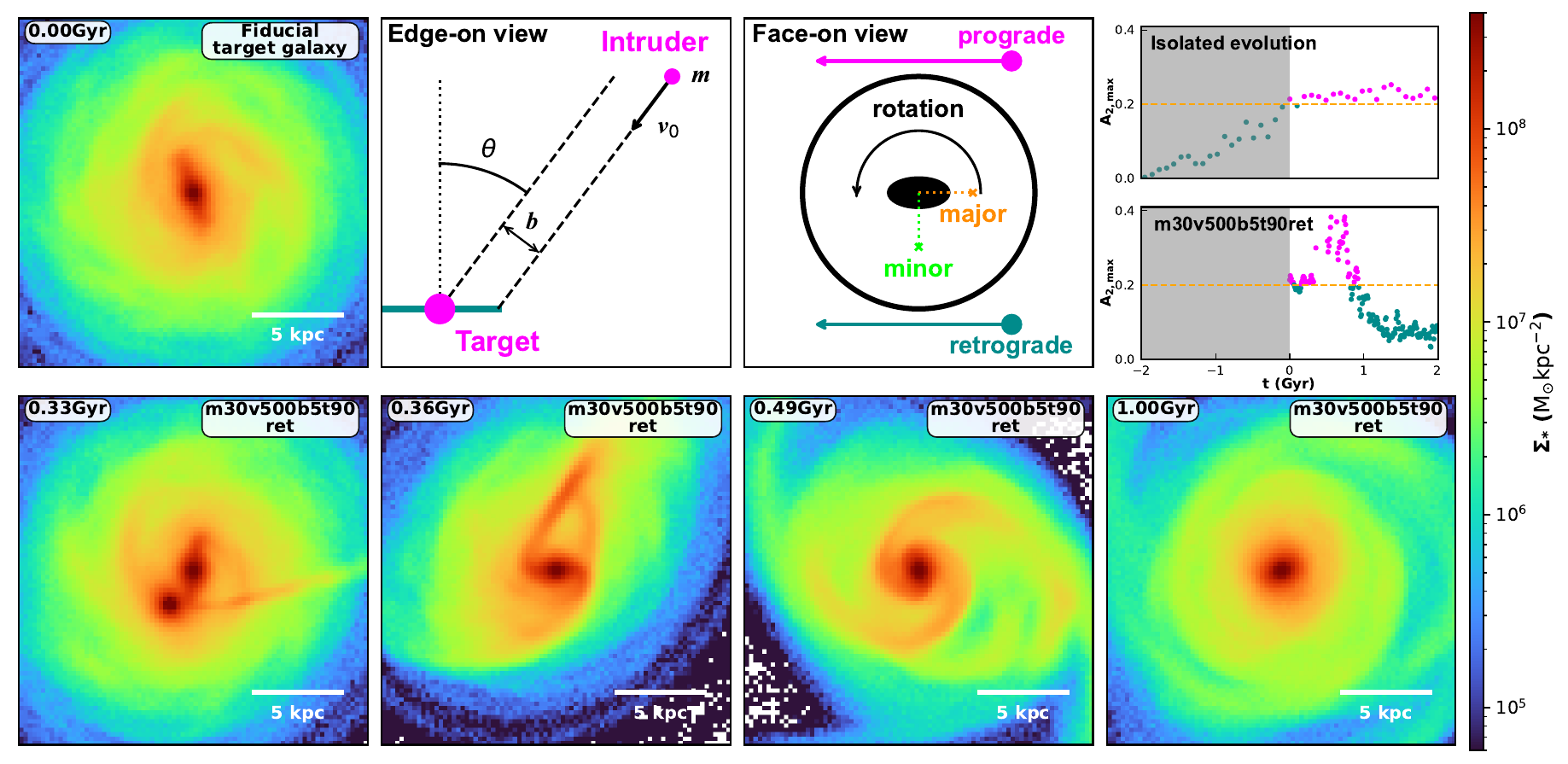}
    \caption{Upper-left panel: face-on stellar surface density map of the fiducial barred target galaxy at 0\,Gyr. Upper middle-left panel: edge-on schematic diagram of the collision simulation, showing $m$, $V_{\rm 0}$, $b$, and $\theta$. Upper middle-right panel: face-on schematic diagram for illustrating some specific scenarios: prograde (retrograde) collisions and major(minor)-axis collisions. Upper-right panel: the bar strength ($A_{\rm 2,max}$) evolution of the target galaxy during isolated evolution or in the collision simulation `m30v500b5t90ret', with the phase before bar formation (-2\,Gyr to 0\,Gyr) marked with gray shading. The horizontal yellow dashed line corresponds to $A_{\rm 2,max}=0.2$, with points above it colored magenta and the remainder in cyan. Lower panels: face-on stellar surface density map at different times in `m30v500b5t90ret'.}
    \label{fig:1}
\end{figure*}

We then let this galaxy evolve in isolation using the \textit{N}-body/SPH code \textsc{gadget-4} \citep{springel2005,springel2021}, incorporating the sub-resolution multiphase interstellar medium model\footnote{The model assumes a thermal instability triggered above a critical density threshold, producing a two-phase medium which comprises cold clouds embedded in a tenuous gas at pressure equilibrium. Stars form within these clouds on a specific timescale, and the resulting supernovae supply an energy to the surrounding gas. We adopt typical values \citep{springel2021} for parameters in the model: the star formation timescale $t_{*}^0=1.5$\,Gyr, the cloud evaporation parameter $A_{\rm 0}=1000$, and the supernova temperature $T_{\rm SN}=10^8$\,K.} \citep{springel2003} to govern cooling, star formation, and supernova feedback. A bar formed from internal instabilities at approximately 0\,Gyr (with the onset of isolated evolution marked as -2\,Gyr to align with the formal collision simulation timeline) and maintained stable strength until at least 2\,Gyr. We select the snapshot at 0\,Gyr as the initial condition for the fiducial target galaxy in collision simulations. The upper-left panel of Figure~\ref{fig:1} shows its face-on stellar surface density map at 0\,Gyr, with the bar strength evolution in the isolated case displayed in the upper-right panel. Furthermore, in addition to the fiducial model with a 30\% disk gas fraction, we construct a model without gas for comparison (see Section~\ref{subsec:nogas}).

The intruder is modeled as a spherical galaxy consisting of an NFW dark matter halo and a Hernquist stellar bulge. Preliminary testing confirmed that intruders with insufficient mass produce negligible effects on the bar. Consequently, our formal simulations adopted intruder masses of $1\times10^{11}$\,${M}_{\odot}$, $3\times10^{11}$\,${M}_{\odot}$, and $5\times10^{11}$\,${M}_{\odot}$ (`intruder mass' refers to its virial mass throughout the paper), corresponding to mass ratios with the target galaxy of 1:15, 1:5, and 1:3, respectively. The parameters for the intermediate-mass intruder (fiducial model) are shown in Table~\ref{tab:1}. For other two intruders, the masses and sizes of their components are proportionally scaled according to the virial mass. Prior to the formal collision simulations, the intruder was also evolved in isolation for 2\,Gyr.

Notably, we insert a particle with mass of $4.3\times10^{6}$\,${M}_{\odot}$ \citep{gillessen2009} at the center of the target galaxy to represent the MW supermassive black hole (SMBH, but SMBH-related physical processes are not considered). It naturally settles at the potential minimum, i.e., the galactic center, facilitating convenient determination of the target's center for subsequent bar analysis. The softening lengths for dark matter, baryons, and black holes in both isolated evolution and collision simulations are 30\,pc, 10\,pc, and 50\,pc, respectively.
\begin{table}
    \centering
    \caption{Fiducial galaxy models. The displayed parameters are virial mass $M_{\rm vir}$, halo concentration $c$, total disk mass $M_{\rm d}$, initial gas fraction in the disk $f_{\rm g}$, disk scale length $r_{\rm s}$, disk scale height $h$, stellar bulge mass $M_{\rm b}$, bulge scale length $a$, dark matter particle mass $m_{\rm DM}$, baryon particle mass $m_{\rm B}$, dark matter particle number $N_{\rm DM}$, stellar disk particle number $N_{\rm d,s}$, gaseous disk particle number $N_{\rm d,g}$, and bulge particle number $N_{\rm b}$, in sequence.}
    \begin{tabular}{ccc}
    \hline
    Parameter & Target & Intruder \\
    \hline
    $M_{\rm vir}$ ($10^{10}$\,${M}_{\odot}$) & 150 & 30 \\
    $c$ & 9.56 & 11.50 \\
    $M_{\rm d}$ ($10^{10}$\,${M}_{\odot}$) & 4.10 & 0.00 \\
    $f_{\rm g}$ & 0.30 & 0.00 \\
    $r_{\rm s}$ (kpc) & 2.20 & - \\
    $h$ (kpc) & 0.44 & - \\
    $M_{\rm b}$ ($10^{10}$\,${M}_{\odot}$) & 0.80 & 1.47 \\
    $a$ (kpc) & 0.40 & 0.49 \\
    $m_{\rm DM}$ ($10^{4}$\,${M}_{\odot}$) & 20 & 20 \\
    $m_{\rm B}$ ($10^{4}$\,${M}_{\odot}$) & 1 & 1 \\
    $N_{\rm DM}$ & 7,255,000 & 1,426,500 \\
    $N_{\rm d,s}$ & 2,870,000 & 0 \\
    $N_{\rm d,g}$ & 1,230,000 & 0 \\
    $N_{\rm b}$ & 800,000 & 1,470,000 \\
    \hline
    \end{tabular}
    \label{tab:1}
\end{table}

\subsection{Simulation Configurations}
\label{subsec:config}
We position the target galaxy's center at the origin, with its disk lying in the $x-y$ plane. The intruder's initial velocity vector intersects the disk plane at a point, and the initial distance from the intruder to this intersection point is 200\,kpc (see Appendix~\ref{sec:a3} for a test using 400\,kpc), approximately the target's virial radius. The collision simulations are run using \textsc{gadget-4} \citep{springel2021} for approximately 2.1\,Gyr, with snapshot outputs at intervals of roughly 10\,Myr. 

The simulation configuration is primarily characterized by four parameters: intruder mass $m$, intruder initial velocity $V_{\rm 0}$, impact parameter $b$, and inclination angle $\theta$, as schematically illustrated in Figure~\ref{fig:1}. In our simulations, $m$ takes three values: $1\times10^{11}$\,${M}_{\odot}$, $3\times10^{11}$\,${M}_{\odot}$, and $5\times10^{11}$\,${M}_{\odot}$, as described in Section~\ref{subsec:models}. We also have three values for $V_{\rm 0}$: 500\,km\,s$^{-1}$, 350\,km\,s$^{-1}$, and 250\,km\,s$^{-1}$, spanning a range from fast flyby speeds in massive clusters \citep[e.g.,][]{carlberg2001} to bound scenarios. The first two only result in a single collision within the simulation duration (2.1\,Gyr), while the third leads to multiple collisions and eventual merger. The possible values of $b$ include 0\,kpc, 5\,kpc, and 10\,kpc, where 0\,kpc corresponds to a central collision and non-zero values represent different extents of off-center scenarios. Finally, the values of $\theta$ contain 0$^{\circ}$ (vertical collision), 45$^{\circ}$, and 90$^{\circ}$. The upper middle-right panel of Figure~\ref{fig:1} illustrates that when both $b$ and $\theta$ are non-zero, collisions can be classified as prograde or retrograde depending on the orientation of the orbital angular momentum relative to the target's disk angular momentum. Additionally, since barred galaxies are non-axisymmetric, even vertical collisions with identical $b$  are geometrically distinct if impact points differ. Configurations where the impact point lies along the bar's major or minor axis (termed major-axis collision and minor-axis collision; see Section~\ref{subsec:point}) are shown in Figure~\ref{fig:1}.

All possible values of parameters in collision simulations are listed in Table~\ref{tab:2}, and the specific values adopted in each simulation are annotated in the label at the upper-right corner of each panel in Figure~\ref{fig:2} and Figure~\ref{fig:4}. For example, `m30v500b5t45pro' denotes a simulation with $m=3\times10^{11}$\,${M}_{\odot}$, $V_{\rm 0}=500$\,km\,s$^{-1}$, $b=5$\,kpc, $\theta=45$$^{\circ}$, and a prograde collision orbit; whereas `m50v500b5t0maj' indicates $m=5\times10^{11}$\,${M}_{\odot}$, $V_{\rm 0}=500$\,km\,s$^{-1}$, $b=5$\,kpc, $\theta=0$$^{\circ}$, and an impact point lying along the bar's major axis. Furthermore, simulations suffixed with `f0' indicate a disk gas fraction $f_{\rm g}=0$, while those suffixed with `mid' (`low') denote medium-resolution (low-resolution) simulations (see Section~\ref{subsec:resolution}).

\begin{table}[]
    \centering
    \caption{Parameter values used in collision simulations. Prograde, retrograde, major-axis, and minor-axis scenarios are represented by `pro', `ret', `maj', and `min', respectively. The `pro/ret' suffix only appears in simulations where $b \neq 0$ and $\theta \neq 0$, while the `maj/min' suffix indicates that the intruder's initial distance has been adjusted to control the impact point (see Section~\ref{subsec:point}). Examples of each parameter in the simulation labels are displayed in magenta in the third column.}
    \begin{tabular}{ccc}
    \hline
    Parameter & Values & Example \\
    \hline
    $m$ ($10^{10}$\,${M}_{\odot}$) & 10, 30, 50 & \textcolor{magenta}{m30}v500b0t0 \\
    $V_{\rm 0}$ (km\,s$^{-1}$) & 250, 350, 500 & m30\textcolor{magenta}{v500}b0t0 \\
    $b$ (kpc) & 0, 5, 10 & m30v500\textcolor{magenta}{b0}t0 \\
    $\theta$ ($^{\circ}$) & 0, 45, 90 & m30v500b0\textcolor{magenta}{t0} \\
    orbital type & default, pro, ret & m30v500b5t90\textcolor{magenta}{pro} \\
    $f_{\rm g}$ & default(0.3), 0.0 & m30v500b0t0\textcolor{magenta}{f0} \\
    impact location & default, maj, min & m30v500b5t0\textcolor{magenta}{maj} \\
    mass resolution & default, mid, low & m30v500b0t0\textcolor{magenta}{low} \\
    \hline
    \end{tabular}
    \label{tab:2}
\end{table}

\subsection{Fourier Decomposition}
\label{subsec:fourier}
We perform Fourier decomposition \citep[e.g.,][]{athanassoula2002} on the face-on stellar surface density of the target galaxy to quantify the bar strength. Specifically, we first construct a cylinder with a radius of 8\,kpc and a height of 10\,kpc (5\,kpc above and below the disk, respectively; the adopted value has nearly no impact on subsequent results within a reasonable range), centered on the disk's center. This cylinder is then subdivided into concentric cylindrical shells with a width of 0.1\,kpc. For a cylindrical shell at radius $R$, $A_{\rm 2}$ is defined as the ratio of the second-order term to the zero-order term of the Fourier expansion:
\begin{equation}
    A_{\rm 2}(R)=\frac{|\Sigma_{\rm j}m_{\rm j}e^{2i\alpha_{\rm j}}|}{\Sigma_{\rm j}m_{\rm j}},
\end{equation}
where $m_{\rm j}$ is the mass of the $j$-th star particle belonging to this shell and $\alpha_{\rm j}$ is its angular coordinate in the disk plane. The maximum value of $A_{\rm 2}(R)$ across all cylindrical shells, denoted as $A_{\rm 2,max}$, is defined as the bar strength. When $A_{\rm 2,max}\geq0.2$, the target galaxy is considered to host a bar \citep[e.g.,][]{jimenez2024}. However, when $\sim 0.1<A_{\rm 2,max}<0.2$, the target may still exhibit weak bar-like structures \citep[e.g.,][]{hebrail2025}, which will be also analyzed (see Section~\ref{sec:results} for details). In addition, the bar length $r$ is defined as the radial extent where $A_{\rm 2}(R)$ drops to 70\% of its peak value \citep[e.g.,][]{ghosh2024}. 

To compute the pattern speed $\Omega_{\rm p}$, we need to calculate the phase of the $m=2$ mode, $\Phi(R)$:
\begin{equation}
    \Phi(R)=\frac{1}{2}\arctan\left[\frac{\Sigma_{\rm j}m_{\rm j}\sin(2\alpha_{\rm j})}{\Sigma_{\rm j}m_{\rm j}\cos(2\alpha_{\rm j})}\right],
\end{equation}
which remains approximately constant within the extent of the bar. Taking the derivative of $\Phi$ with respect to time yields $\Omega_{\rm p}$.

The above method may have limitations in interacting galaxies, because spurious classification of tidal features as bars can lead to unphysical fluctuations in the lengths. We therefore implement a phase consistency check: if the phase at the derived radius differs from the phase in the inner region by more than 15$^{\circ}$, we consider the result unreliable and instead adopt the largest radius over which the phase remains consistent as the bar length. However, in cases where the disk is strongly deformed, periodic alignment and misalignment between tidal structures and the bar can still lead to fluctuations.

\section{Results}
\label{sec:results}

\begin{figure*}
    \centering
    \includegraphics[width=0.9\linewidth]{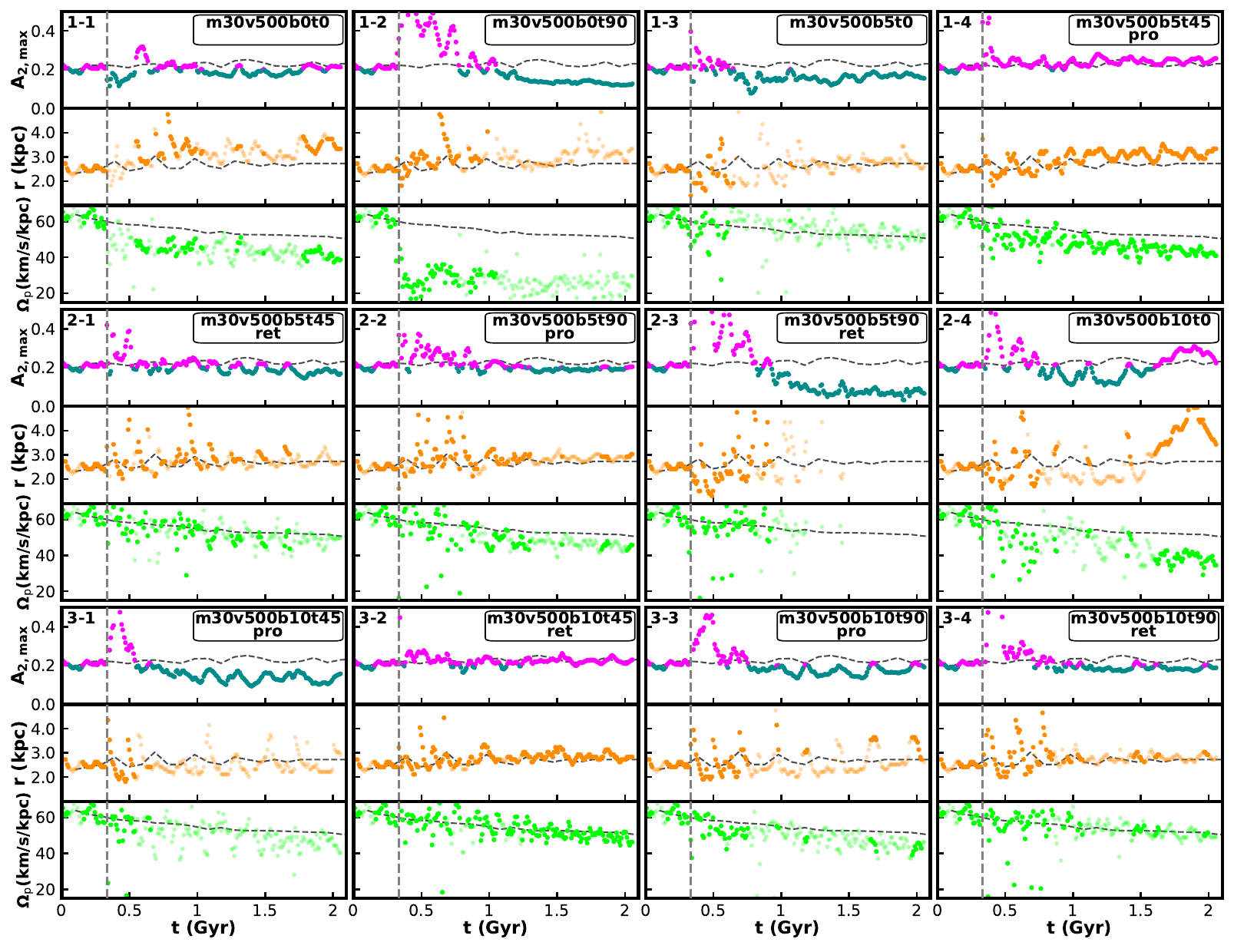}
    \caption{The evolution of bar strength $A_{\rm 2,max}$, length $r$, and pattern speed $\Omega_{\rm p}$ over time for 12 fiducial collision simulations. The upper-left corner of each panel is marked with row and column indices for easy navigation. The upper-right corner displays the simulation label, indicating the values of $m$, $V_{\rm 0}$, $b$, and $\theta$. Prograde and retrograde simulations are annotated with `pro' and `ret', respectively. The `pro/ret' suffix only appears in simulations where $b \neq 0$. In the $A_{\rm 2,max}$ sub-panel, points above the bar existence threshold of 0.2 are colored magenta, while the rest are colored cyan. For times where $A_{\rm 2,max}>0.1$, the values of $r$ and $\Omega_{\rm p}$ are calculated; but if $A_{\rm 2,max}<0.2$, $r$ and $\Omega_{\rm p}$ points are rendered semi-transparently. The gray vertical dashed line marks the time of the collision, which is the single collision event for all simulations. The black dashed line shows the evolution of bar parameters in the isolated simulation (see Section~\ref{subsec:models}).}
    \label{fig:2}
\end{figure*}

\subsection{Overview of Simulation Results}
\label{subsec:3.1}
At the start of each simulation, the intruder moves toward the target galaxy and collides after several hundred million years, with the exact timing dependent on the values of $m$, $V_{\rm 0}$, and $b$. Due to variations in the interaction parameters, the bar in different simulations experiences different degrees of influence. Figure~\ref{fig:2} shows the evolution of bar strength $A_{\rm 2,max}$, length $r$, and pattern speed $\Omega_{\rm p}$ over time for 12 fiducial collision simulations (see Figure~\ref{fig:S2} for face-on stellar surface density of the target galaxy at the end of each simulation). These 12 simulations have the same intruder mass and initial velocity, i.e., $m=3\times10^{11}$\,${M}_{\odot}$, and $V_{\rm 0}=500$\,km\,s$^{-1}$. Figure~\ref{fig:2} displays these simulations in order of increasing $b$ value, and simulations with the same $b$ are shown in order of increasing $\theta$. The results of another 12 simulations with $m=3\times10^{11}$\,${M}_{\odot}$, $V_{\rm 0}=350$\,km\,s$^{-1}$ and another 12 simulations with $m=5\times10^{11}$\,${M}_{\odot}$, $V_{\rm 0}=500$\,km\,s$^{-1}$, which exhibit similar outcomes to the 12 fiducial simulations, are shown in Figure~\ref{fig:S1}. Additionally, some special simulations discussed in Section~\ref{sec:discussions} are shown in Figure~\ref{fig:4}.

For most simulations, bar strength exhibits a rapid increase immediately following the collision\footnote{Considering that we use a SMBH to mark the galactic center, the temporary displacement of the SMBH after collision could also affect bar strength measurements. However, this mechanism has minimal impact here. For instance, the peak differences between the SMBH position and the target's center of mass in `m30v500b0t0' and `m30v500b0t90' are similar (0.2\,kpc and 0.25\,kpc, respectively), yet the bar strength changes at $\sim$0.4\,Gyr in Figure~\ref{fig:2} show opposite trends.} (marked by vertical gray dashed lines in Figure~\ref{fig:2}). This is due to the geometrically non-axisymmetric collision configuration, which induces bar-like structures and enhances bar strength, as also observed in \citet{zhou2025}. In high-inclination scenarios, such intense structures can even become dominant, leading to chaotic $r$ and $\Omega_{\rm p}$ values that overflow the ordinate axis range in Figure~\ref{fig:2}, as seen in `m30v500b5t90ret' (panel 2-3). In contrast, central vertical impacts ($b=0$\,kpc and $\theta=0$$^{\circ}$) do not produce a rapid increase in bar strength; instead, they cause an immediate slight reduction. However, these changes in bar strength are transient and quickly recover. The long-term effects of collisions on bars are analyzed below.

\subsection{High-Inclination Configurations Can Significantly Affect the Bar}
\label{subsec:3.2}
The first result summarized from the simulations is that bars are highly robust \citep[e.g.,][]{athanassoula1999,shen2004} in MW-like galaxies. Under the vast majority of interaction parameters, bars are not significantly disrupted.

Nevertheless, bars are not indestructible. We note that off-center high-inclination retrograde collisions can effectively destroy bars (see lower panels of Figure~\ref{fig:1} for an example), as demonstrated by `m30v500b5t90ret' (panel 2-3) in Figure~\ref{fig:2}. In such collision configurations, the orbital angular momentum of the intruder is antiparallel to the bar's rotation angular momentum, enabling the bar to dissolve. However, if $b$ increases from 5\,kpc to 10\,kpc, the disruptive effect is reduced, as seen in `m30v500b10t90ret' (panel 3-4). Therefore, the magnitude of off-centering should be moderate.

Several other simulations also disrupted bars notably, such as `m30v500b0t90'(panel 1-2). However, the performance of these collision configurations is inconsistent, as their corresponding sibling simulations (with identical $b$ and $\theta$ but differing $m$ or $V_{\rm 0}$), namely `m30v350b0t90' (panel 1-2 of Figure~\ref{fig:S1}), fail to significantly weaken bars. Due to differences in mass and initial velocity, the intruder's arrival time at the disk plane varies, leading to differing impact points relative to the bar, which may account for the discrepancies described above, with detailed discussion provided in Section~\ref{subsec:point}.

Pattern speed and length are also worthwhile properties to investigate if a bar is present, i.e., when $A_{\rm 2,max}\geq 0.2$. However, even when bar strength slightly falls below the conventional threshold of 0.2 for bar existence, weak bar-like structures remain identifiable in the simulations. Therefore, we calculate bar pattern speed and length for all the times with $A_{\rm 2,max}>0.1$ using the method described in Section~\ref{subsec:fourier}, but scatter points for $0.1<A_{\rm 2,max}<0.2$ are rendered semi-transparent in Figure~\ref{fig:2}.

If the target galaxy evolves in isolation, its bar length gradually increases over time due to growth, while pattern speed gradually decreases due to effective outward angular momentum transport around resonance regions \citep[e.g.,][]{sellwood1980,debattista2000,jimenez2024}, as shown by the black dashed lines in Figure~\ref{fig:2}. For most collision simulations, the natural deceleration of bars remains largely unaffected, with pattern speeds decreasing from $>60$\,km\,s$^{-1}$\,kpc$^{-1}$ to $40–50$\,km\,s$^{-1}$\,kpc$^{-1}$ within $\sim$2.1\,Gyr.

However, central high-inclination collisions can substantially reduce the bar's pattern speed to $<20$\,km\,s$^{-1}$\,kpc$^{-1}$ within a few hundred million years, as demonstrated in `m30v500b0t90' (panel 1-2). Some other simulation studies, such as \citet{jimenez2025}, also show that specific interactions can significantly slow down the bar, even temporarily stopping it. For collision simulations where pattern speed almost remains unaffected, slight variations may still exist across different runs, likely also related to differences in impact points, as detailed in Section~\ref{subsec:point}. Furthermore, although bar length exhibits fluctuations, a negative correlation generally exists between the bar length and the pattern speed at the end of simulations, modulated by angular momentum conservation.

\subsection{Influence of Individual Parameters}
\label{subsec:3.3}
\begin{figure*}
    \centering
    \includegraphics[width=0.7\linewidth]{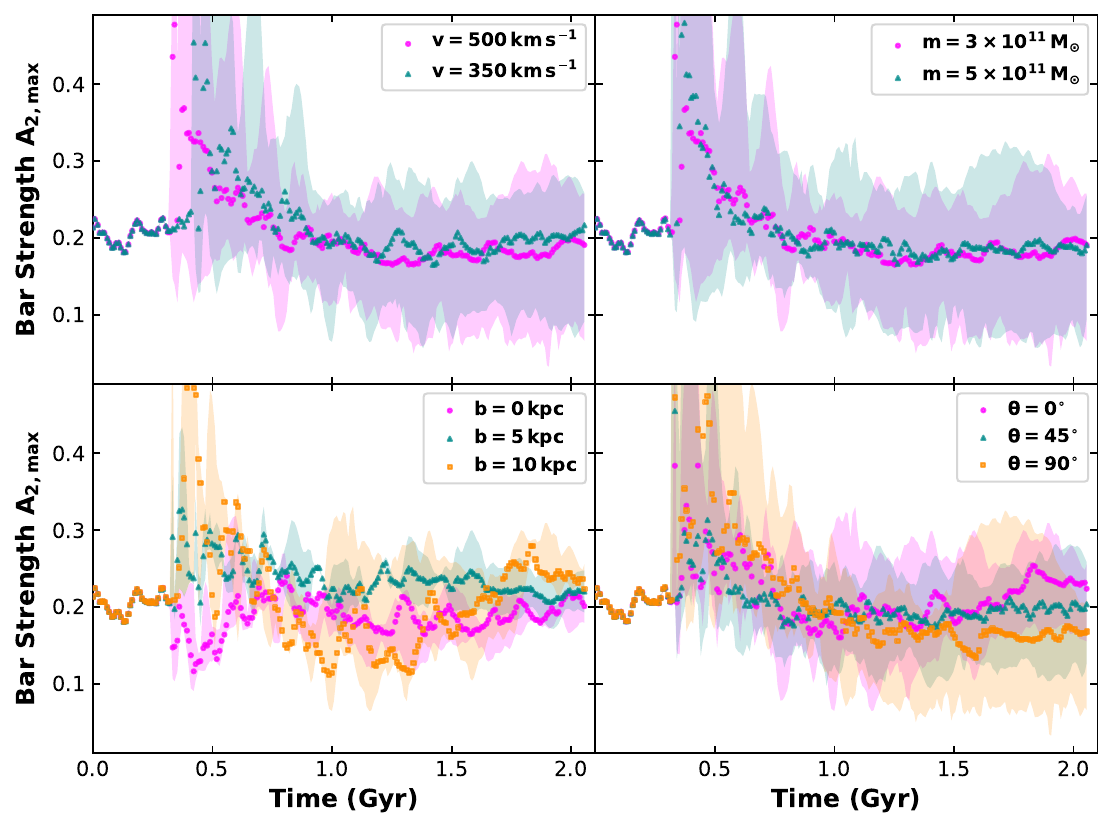}
    \caption{Evolution of the median bar strength over time for simulations sharing a specific parameter value in Figure~\ref{fig:2} and Figure~\ref{fig:S1}. The shaded region behind each median scatter point indicates the range from the minimum to the maximum value.}
    \label{fig:3}
\end{figure*}
In simulations with different parameters, bars can be effectively destroyed, mildly weakened, or remain entirely unaffected. To investigate the general influence of each parameter, we classify, integrate, and compare the results of the 12 fiducial simulations (Figure~\ref{fig:2}) and the 24 sibling simulations (Figure~\ref{fig:S1}).

\subsubsection{Higher Velocities Increase Bar Destruction Efficiency}
First, the influence of $V_{\rm 0}$ is analyzed. The twelve simulations in Figure~\ref{fig:2} are treated as an ensemble. For each time, the median bar strength across these simulations is plotted as a magenta point in the upper-left panel of Figure~\ref{fig:3}, with the background magenta shaded region indicating the range from minimum to maximum values. The twelve simulations from row~1 to row~3 of Figure~\ref{fig:S1} are treated as another ensemble, plotted in cyan. We find that after the collision, the $A_{\rm 2,max}$ values for $V_{\rm 0}=350$\,km\,s$^{-1}$ are slightly higher than those for $V_{\rm 0}=500$\,km\,s$^{-1}$, though the difference diminishes over the final $\sim$0.5\,Gyr. This suggests that higher relative velocities are more conducive to bar destruction. However, if the velocity is too low, the intruder not only collides with the target galaxy but also merges with it. In merging cases, the bar can be severely destroyed due to central mass concentration \citep[e.g.,][]{athanassoula1999}, even in collision configurations that typically do not disrupt bars, such as central vertical impacts (`m30v250b0t0').

\subsubsection{Intruder Mass Has Negligible Impact on Bar Stability}
The upper-right panel of Figure~\ref{fig:3} compares the ensembles for $m=5\times10^{11}$\,${M}_{\odot}$ and $m=3\times10^{11}$\,${M}_{\odot}$ (mass ratio of 1:3 and 1:5), with plotting data derived from Figure~\ref{fig:2} and row~4 to row~6 of Figure~\ref{fig:S1}, respectively. They show nearly no difference, indicating that intruder mass does not play a primary role in collision-induced bar destruction. This conclusion holds only within the mass range studied, in which the bar of the target galaxy is not easily destroyed. For more massive intruders, the bar's response to collisions may exhibit significant mass-dependent effects. Moreover, the mass cannot be too small either; otherwise, even the configuration most effective at destroying bars (i.e. the off-center high-inclination retrograde collision `m10v500b5t90ret') fails to be effective.

\subsubsection{Retrograde Collisions Induce Stronger Bar Weakening}
Caution should be exercised when analyzing the influence of $b$, as simulations with $b\neq0$\,kpc are far more complex than those with $b=0$\,kpc due to prograde or retrograde characteristics. Considering that parameter coupling prevents direct comparison of some simulations, here we only compare vertical collision scenarios that lack prograde or retrograde characteristics. Therefore, the sample size is relatively limited, and the following conclusion requires further validation through subsequent work. We use `m30v500b0t0', `m30v350b0t0', `m50v500b0t0'; `m30v500b5t0', `m30v350b5t0', `m50v500b5t0'; and `m30v500b10t0', `m30v350b10t0', `m50v500b10t0' in Figure~\ref{fig:2} and Figure~\ref{fig:S1} to represent scenarios with $b$ of 0\,kpc, 5\,kpc, and 10\,kpc, respectively. The lower-left panel of Figure~\ref{fig:3} shows that, considering $b$ alone, off-center impacts result in higher bar strength in the end, consistent with \citet{zhou2025}. However, as discussed in Section~\ref{subsec:3.2}, once the collision becomes retrograde, the result changes entirely.

\subsubsection{Higher Inclination angles Cause Greater Bar Disruption}
Similarly, considering that some simulations are not directly comparable, we exclude simulations with $b=0$\,kpc when studying the effect of $\theta$. The remaining 30 simulations in Figure~\ref{fig:2} and Figure~\ref{fig:S1} are grouped by $\theta$ and plotted in the lower-right panel of Figure~\ref{fig:3}. Curves with higher $\theta$ values occupy lower positions at the simulation endpoint, indicating that bar disruption is more likely in high-inclination collision simulations, reaffirming the conclusion in Section~\ref{subsec:3.2}. However, within a few hundred million years after the collision, the curve for $\theta=90$$^{\circ}$ is actually the highest, while the curve for $\theta=0$$^{\circ}$ lies lower. These are transient features arising from the collision geometry itself, as discussed in Section~\ref{subsec:3.1}, and dissipate over time.

\section{Discussions}
\label{sec:discussions}

\begin{figure*}
    \centering
    \includegraphics[width=0.9\linewidth]{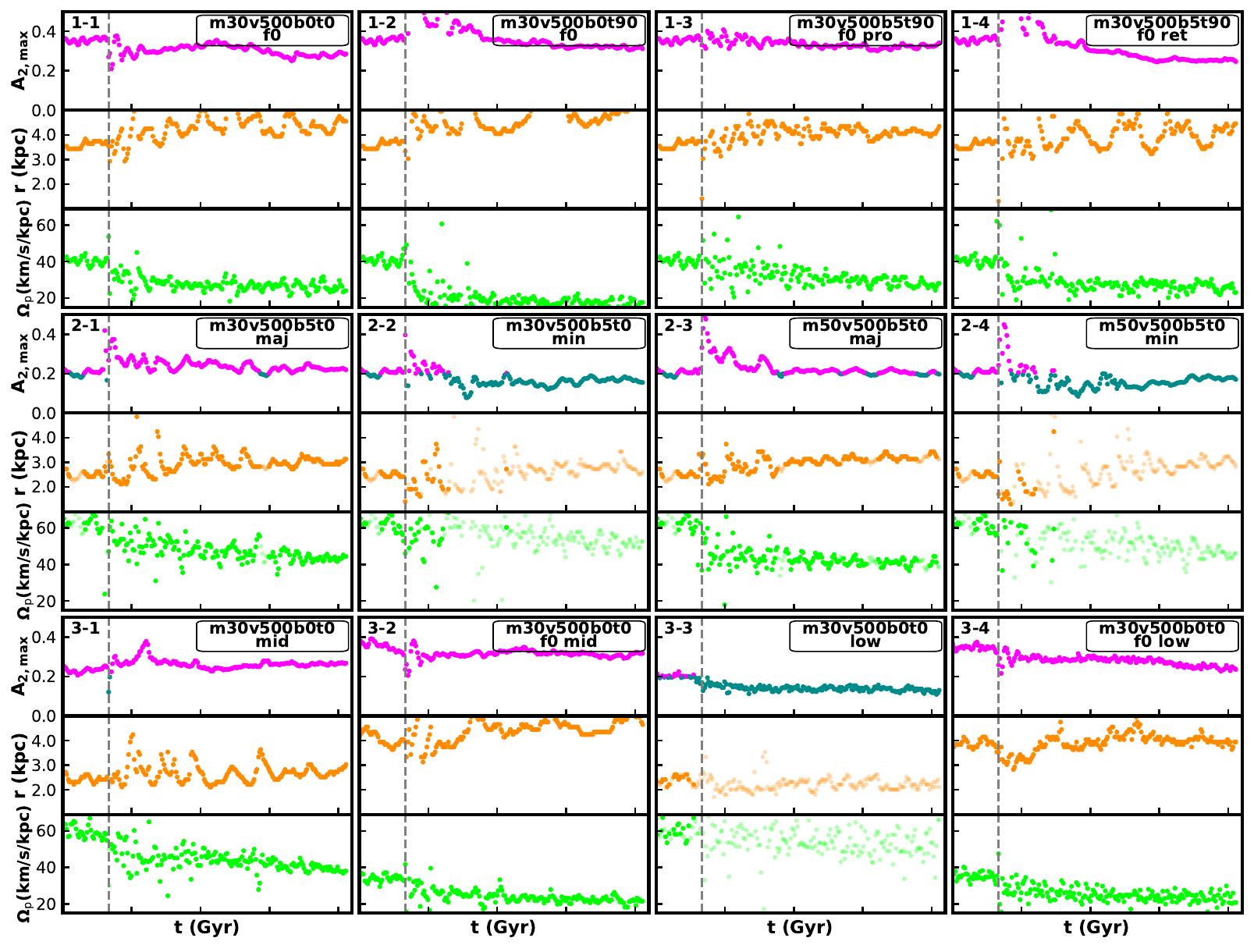}
    \caption{The evolution of bar strength $A_{\rm 2,max}$, length $r$, and pattern speed $\Omega_{\rm p}$ over time for the 12 collision simulations mentioned in Section~\ref{sec:discussions}. Prograde, retrograde, major-axis, minor-axis, gas-free, medium-resolution, and low-resolution simulations are annotated with `pro', `ret', `maj', `min', `f0', `mid', and `low', respectively. Since there is no need to compare these 12 simulations with the isolated scenario, the evolution of the bar parameters in the isolated collision is not shown, unlike in Figure~\ref{fig:2}.}
    \label{fig:4}
\end{figure*}

\subsection{Simulations Without Gas}
\label{subsec:nogas}
The presence or absence of gas components can significantly influence the properties of barred galaxies \citep[e.g.,][]{athanassoula2013}, with \citet{zhou2025} also demonstrating that bars form more readily via collisions in gas-free environments. To investigate the role of gas in collision-induced bar destruction, we decide to conduct several simulations without gas components.

Compared to the fiducial target galaxy model in Table~\ref{tab:1}, we set the disk gas fraction to $f_{\rm g}=0$ while keeping all other parameters unchanged. After 2\,Gyr of isolated evolution, a bar that is stronger but slower than in the gas-included case spontaneously forms. We then introduce the intruder and conduct collision simulations with varying parameters (four representative ones are shown in the first row of Figure~\ref{fig:4}). Although bars here are difficult to completely destroy due to their initially high strength, they can be weakened after collisions.

Consistent with the conclusions in Section~\ref{subsec:3.2}, only off-center high-inclination retrograde collisions can effectively weaken bars, and only central high-inclination collisions can drastically slow the pattern speed. Therefore, gas does not play a significant role in this scenario. The disruption of bars by collisions primarily stems from gravitational interactions rather than gas-related physical processes.

\subsection{Influence of Impact Point Relative to the Bar}
\label{subsec:point}
Due to the non-axisymmetric bar in the target galaxy, even with identical $b$ and $\theta$, differences in impact points result in geometrically inequivalent configurations. \textit{N}-body simulations by \citet{nishida1996} on off-center vertical collisions indicate that an intruder passing at a distance from the galactic center along the bar's minor axis can disrupt the bar, whereas passage along the major axis cannot (schematic illustrations of major-axis and minor-axis collisions are shown in Figure~\ref{fig:1}). \textit{N}-body/SPH simulations by \citet{berentzen2003} for these two scenarios also yielded similar results.

To validate the influence of collision points, we conduct four simulations (the second row of Figure~\ref{fig:4}). As described in Section~\ref{subsec:config}, for `m30v500b5t0' (panel 1-3 in Figure~\ref{fig:2}), the intruder's initial distance to the disk plane is 200\,kpc. If we want to control the impact point location, we can finely adjust this distance to alter its arrival time at the disk plane. When the initial distance is set to 185\,kpc and 200\,kpc (coincidentally exactly 200\,kpc), the intruder impacts the disk plane along the bar's major axis and minor axis, respectively.

As shown in Figure~\ref{fig:4}, the final $A_{\rm 2,max}$ value in `m30v500b5t0min' (panel 2-2) is lower than that in `m30v500b5t0maj' (panel 2-1), and the results for `m50v500b5t0min' and `m50v500b5t0maj' (panel 2-4 and 2-3) are consistent with this trend. Therefore, minor-axis collisions indeed exert a stronger weakening effect on bars than major-axis collisions. The underlying mechanism may be that off-center impacts along the major axis cannot induce significant displacement of the bar's center, as bar mass is concentrated along this axis. In contrast, minor-axis collisions generate strong one-armed perturbations, displacing the central region of the bar structure and mixing it with the galaxy's rotational motion, ultimately leading to bar disruption \citep{nishida1996}. Furthermore, the pattern speed of the target galaxy after a minor-axis collision is also higher than that after a major-axis collision.

Upon inspection, differences in the evolutionary trends of $A_{\rm 2,max}$ and $\Omega_{\rm p}$ among some sibling simulations (with identical $b$ and $\theta$, but different $V_{\rm 0}$ or $m$) in Figure~\ref{fig:2} and Figure~\ref{fig:S1} are partly caused by variations in the impact point relative to the bar, due to differences in the intruder's arrival time at the disk plane.

\subsection{Influence of Mass Resolution}
\label{subsec:resolution}
In our simulations, central vertical impacts result in bar strength remaining nearly unchanged from the initial value. However, \citet{berentzen2003} utilized \textit{N}-body/SPH simulations to demonstrate that central vertical collisions can severely destroy bars, proving more effective than major-axis or minor-axis vertical collisions. Considering that \citet{berentzen2003} adopted mass resolutions of $2.4\times10^{6}$\,${M}_{\odot}$ for stars and $9.6\times10^{5}$\,${M}_{\odot}$ for gas, differing from our resolution ($1\times10^{4}$\,${M}_{\odot}$), we conduct two low-resolution runs with baryon particle mass of $1\times10^{6}$\,${M}_{\odot}$ and two medium-resolution runs with baryon particle mass of $1\times10^{5}$\,${M}_{\odot}$ (four simulations in the third row of Figure~\ref{fig:4}).

Comparing `m30v500b0t0', `m30v500b0t0mid', and `m30v500b0t0low', we find that bars in simulations with high and medium resolutions are both robust, but the bar is indeed more easily disrupted in the low-resolution run. This indicates that within a certain range, variations in mass resolution do not significantly affect the stability of the bar. However, if the resolution is too low to resolve the non-circular orbits of the particles that constitute the bar, the bar becomes susceptible to dissolution. Therefore, when studying bar destruction, the resolution should be better than $\sim$ $1\times10^{5}$\,${M}_{\odot}$.

Additionally, although the resolution of 2-dimensional \textit{N}-body simulations in \citet{nishida1996} was also low, their central vertical impacts still failed to destroy bars. This may be because their simulations did not include gas, making bars intrinsically stronger and more resistant to complete disruption, as illustrated by `m30v500b0t0f0low' (panel 3-4) in Figure~\ref{fig:4}.

\section{Summary}
\label{sec:summary}
In this work, we conduct a detailed investigation of the disruptive effects of galactic collisions (non-merging scenarios) on stellar bars hosted by MW-like galaxies. A gas-included MW-like target galaxy is constructed, which spontaneously forms a stable bar after 2\,Gyr of isolated evolution. We then introduce an \textit{N}-body spherical intruder that collides with the target disk, conducting multiple simulations by varying intruder mass $m$, initial velocity $V_{\rm 0}$, impact parameter $b$, inclination angle $\theta$, prograde/retrograde characteristics, and so on. These \textit{N}-body/SPH simulations are run for $\sim$2.1\,Gyr using \textsc{gadget-4} \citep{springel2021}, with a baryon mass resolution of $1\times10^{4}$\,${M}_{\odot}$.

The main conclusions drawn from our simulations are as follows:

1. Although bars are highly robust in MW-like galaxies, they can be effectively disrupted by off-center high-inclination retrograde collisions and significantly decelerated by central high-inclination collisions.

2. Geometric parameters (impact parameter $b$, inclination angle $\theta$, prograde/retrograde) exert a stronger influence on the outcomes, while energy parameters (intruder mass $m$ and initial velocity $V_{\rm 0}$) have relatively minor effects within a certain range.

3. In pure \textit{N}-body tests (without gas components), the configurations that effectively weaken bars are similar to those in fiducial gas-included simulations, indicating that collision-induced bar disruption primarily arises from gravitational forces rather than gas-related processes.

4. Differences in impact locations with respect to the bar major axis also influence outcomes. Compared to impacts aligned with the bar's major axis, those along the minor axis cause greater weakening of the bar but also slow its natural deceleration.

5. Within a certain range, changes in mass resolution do not significantly affect the results. However, the discrepancies between some previous studies and our results may be due to their resolution being too low. When studying the problem of bar destruction, the mass resolution should be better than $\sim$ $1\times10^{5}$\,${M}_{\odot}$.

Our work systematically investigates collisions as a potential mechanism for bar destruction, exploring the parameter space. Previous limited studies are significantly supplemented, and certain contradictory findings are analyzed and explained. Given that collisions are more frequent at high redshifts than in the local universe, this mechanism may partly contribute to the lower bar fraction observed at high redshifts, though its precise contribution requires further simulation and observational studies to determine. Additionally, this work focuses on bar destruction in MW-like galaxies under generalized scenarios; the effects of specific events such as the Gaia-Sausage-Enceladus merger are discussed in an independent work by \citet{chen2026}. 

\begin{acknowledgments}
We thank Juntai Shen, Binhui Chen, and Hui Li for helpful discussions. We thank the anonymous referee for their helpful comments and suggestions. This work is supported by the National Natural Science Foundation of China (grant No. 12225302), the National Key Research and Development Program of China (grant No. 2022YFF0503402), the China Manned Space Program with grant No. CMS-CSST-2025-A10, and the Fundamental Research Funds
for the Central Universities (grant KG202502). \'O. Jim\'enez-Arranz acknowledges funding from ``Swedish National Space Agency 2023-00154 David Hobbs The GaiaNIR Mission'' and ``Swedish National Space Agency 2023-00137 David Hobbs The Extended Gaia Mission''. S. Roca-F\`abrega acknowledges financial support from the Spanish Ministry of Science and Innovation through the research grant PID2021-123417OB-I00, funded by MCIN/AEI/10.13039/501100011033/FEDER, EU.
\end{acknowledgments}

\software{GADGET-4 \citep{springel2005,springel2021}}


\appendix

\section{Appendix Figures}

\begin{figure}
    \centering
    \includegraphics[width=0.8\linewidth]{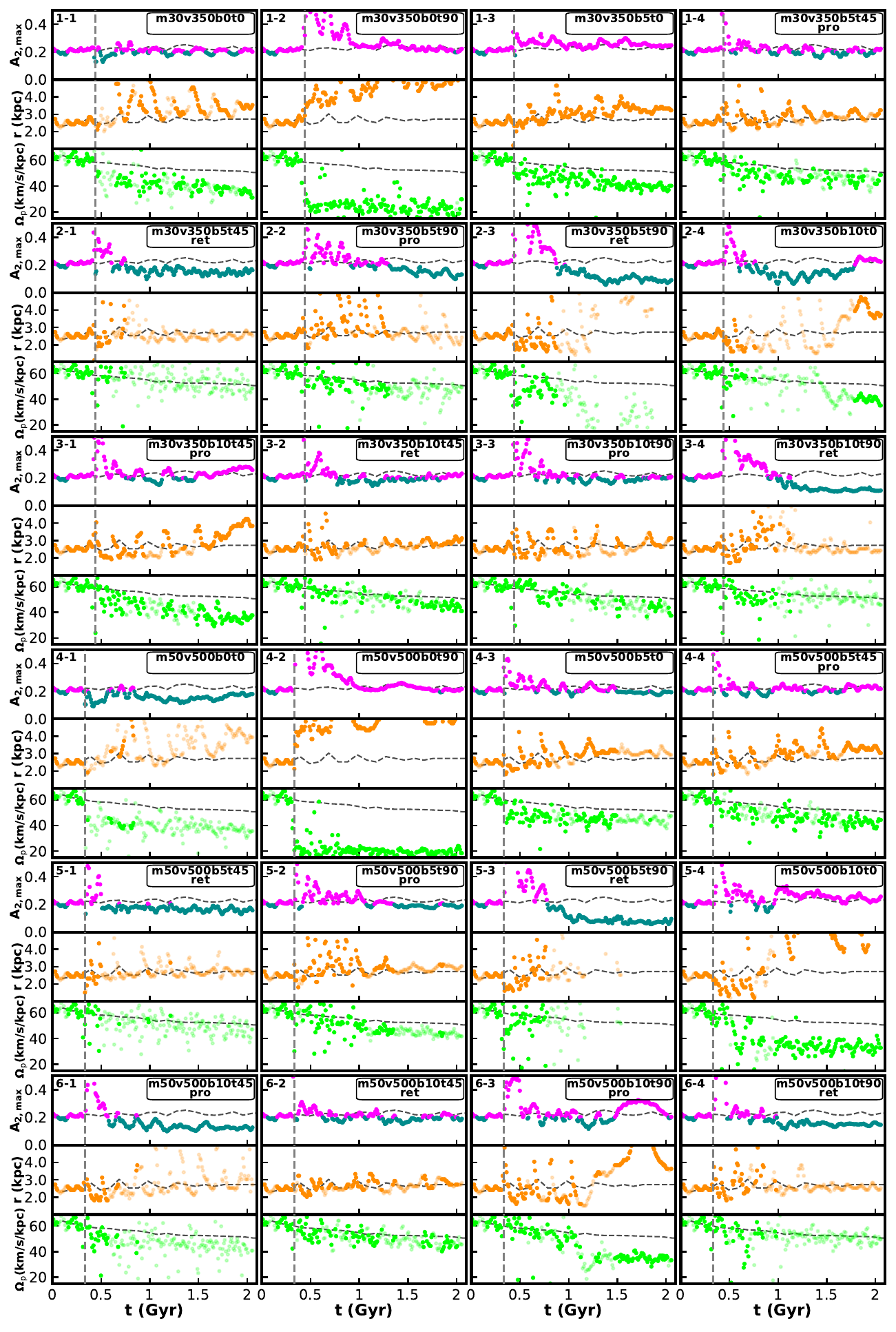}
    \caption{The evolution of bar strength $A_{\rm 2,max}$, length $r$, and pattern speed $\Omega_{\rm p}$ over time for the 24 sibling simulations of the simulations shown in Figure~\ref{fig:2}. The meaning of each element in this figure is the same as in Figure~\ref{fig:2}.}
    \label{fig:S1}
\end{figure}

\begin{figure}
    \centering
    \includegraphics[width=0.75\linewidth]{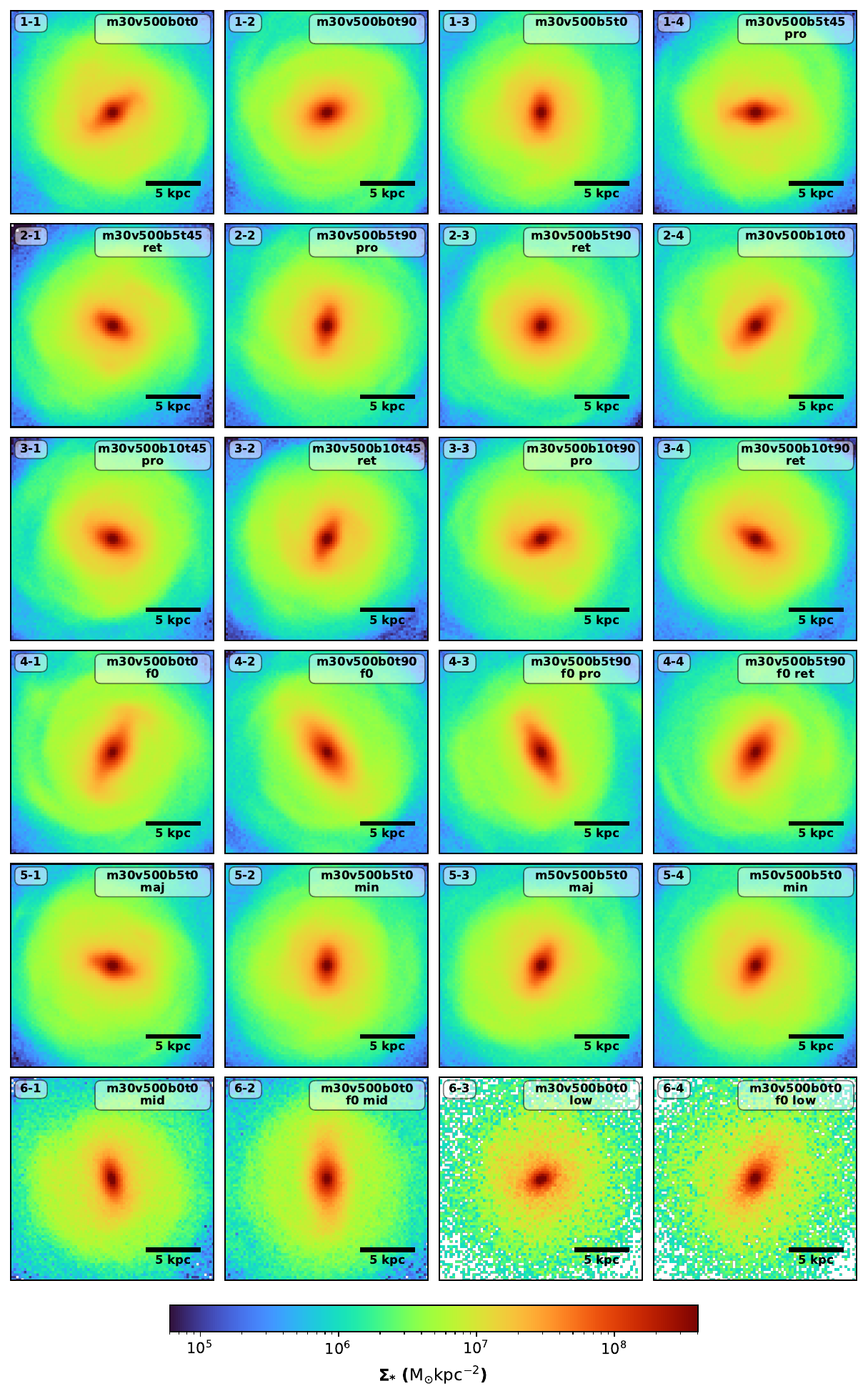}
    \caption{Face-on stellar surface density of the target galaxy at $\sim$2.1\,Gyr for collision simulations in Figure~\ref{fig:2}, Figure~\ref{fig:4} and Figure~\ref{fig:S1}.}
    \label{fig:S2}
\end{figure}
\begin{figure}
    \centering
    \includegraphics[width=0.75\linewidth]{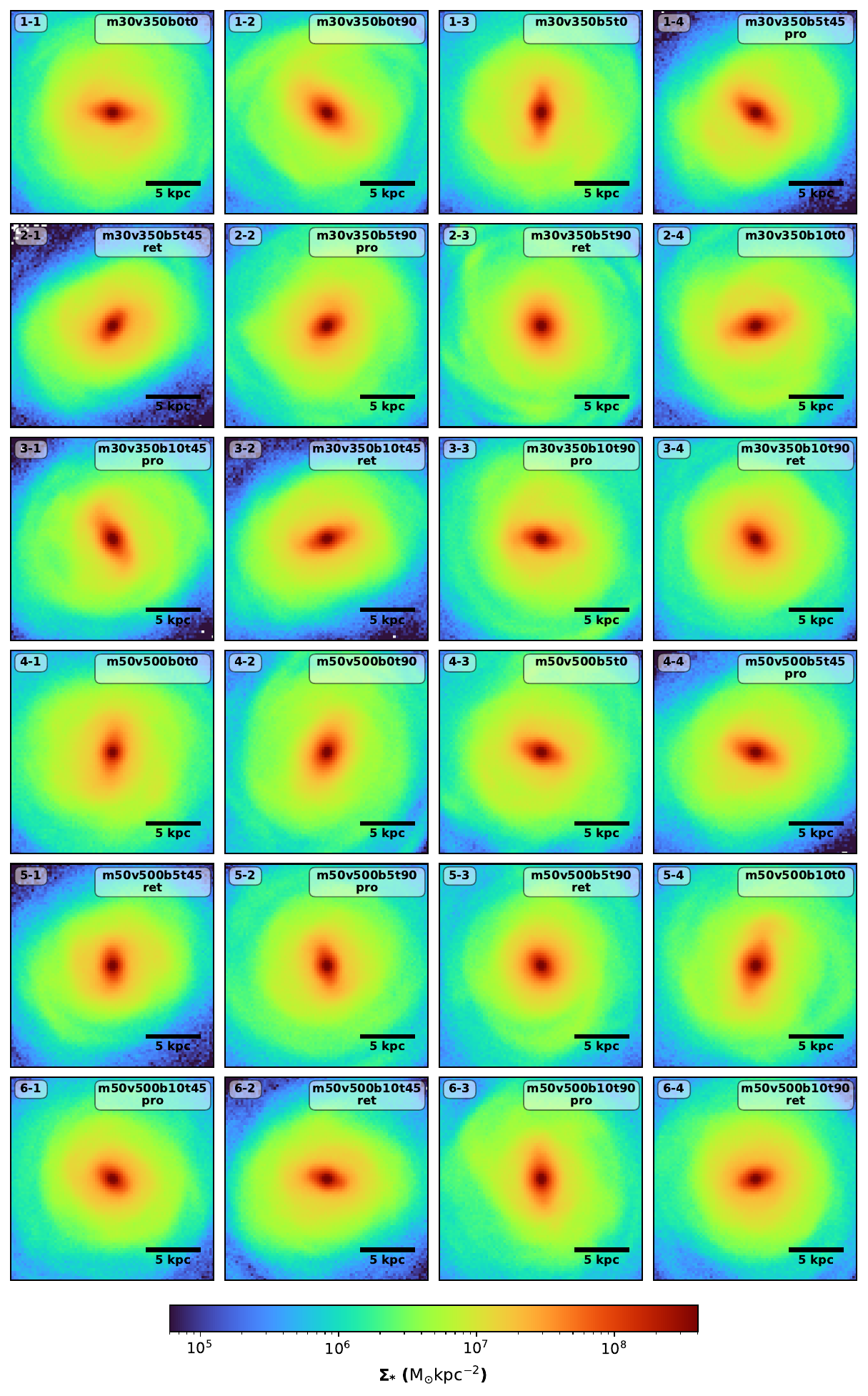}
    \addtocounter{figure}{-1} 
    \caption{(Continued)}
    \label{fig:S2-cont}
\end{figure}

\section{Initial Distance Setting Between Intruder and Target}
\label{sec:a3}
\begin{figure}
    \centering
    \includegraphics[width=0.5\linewidth]{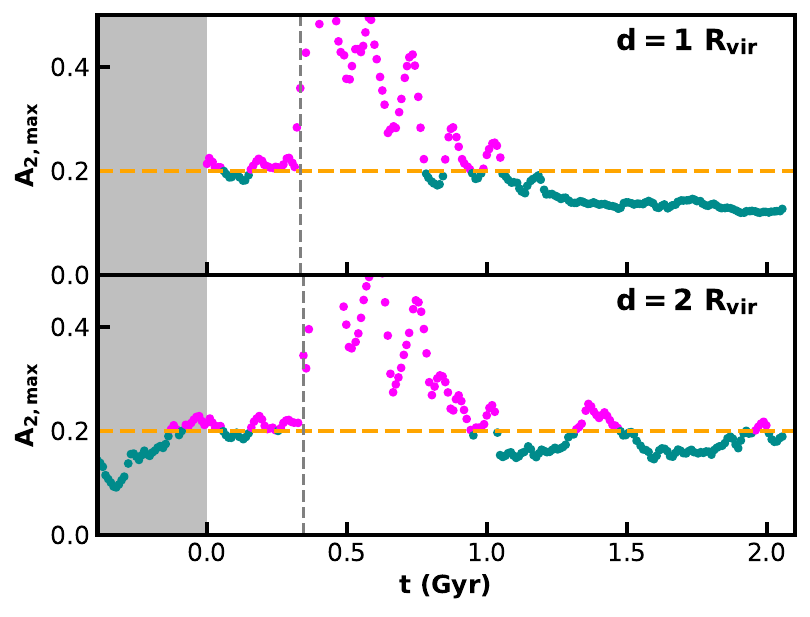}
    \caption{Bar strength evolution for `m30v500b5t90' with the intruder's initial distance to the disk plane set to one and two virial radii. In the second simulation, the intruder took $\sim$0.4\,Gyr (gray shaded region) to travel from 2$R_{\rm vir}$ to 1$R_{\rm vir}$. To align timelines with the first simulation, the start time of the second one is recorded as -0.4\,Gyr.}
    \label{fig:S3}
\end{figure}

As described in Section~\ref{subsec:config}, the initial distance from the intruder to the disk in the simulations is set to approximately the virial radius of the target galaxy (200\,kpc). Considering that tidal forces can be felt by the target's dark matter halo and propagate inward, the bar may experience some influence even before the intruder impacts the disk plane. To ensure that the duration before the collision does not significantly affect the outcomes, we conduct a comparative simulation using `m30v500b0t90' as an example, placing the intruder initially at approximately twice the target galaxy's virial radius (400\,kpc). Prior to this simulation run, the target is evolved in isolation for only 1.6\,Gyr instead of 2\,Gyr, and the intruder's initial velocity is adjusted to 466.64\,km\,s$^{-1}$ instead of 500\,km\,s$^{-1}$, ensuring that the states of the target and intruder at impact closely resemble those in the original simulation. The evolutionary trends of bar strength in these two simulations are similar (see Figure~\ref{fig:S3}), with differences in the final phase likely arising from slight variations in impact timing or stochastic effects. Therefore, the intruder does not significantly affect the bar during the period before impact, confirming that setting the initial distance to one virial radius is appropriate.

\bibliography{bar}{}
\bibliographystyle{aasjournalv7}



\end{document}